\documentclass[prb,aps,showpacs,amsmath,amssymb,preprintnumbers,superscriptaddress,floats,floatfix,twocolumn]{revtex4}
\usepackage[dvips]{graphicx}
\usepackage{dcolumn}
\usepackage{bm}
\usepackage[T1]{fontenc}
\usepackage{t1enc}
\usepackage{textcomp}
\usepackage{indentfirst}

\begin{document}

\title{Magnetization Saturation Process in the Magnonic Anti-dot Structures Based on (Ga,Mn)As: A Magnetometric Study}

\author{K. Dziatkowski}\email{konrad.dziatkowski@fuw.edu.pl}
\affiliation{Faculty of Physics, University of Warsaw, 00-681 Warsaw, Poland}

\author{W. P. Staszewski}
\affiliation{Faculty of Physics, University of Warsaw, 00-681 Warsaw, Poland}

\author{R. Bo\.zek}
\affiliation{Faculty of Physics, University of Warsaw, 00-681 Warsaw, Poland}

\author{J. Szczytko}
\affiliation{Faculty of Physics, University of Warsaw, 00-681 Warsaw, Poland}

\author{J. Gosk}
\affiliation{Faculty of Physics, University of Warsaw, 00-681 Warsaw, Poland}

\author{A. Twardowski}
\affiliation{Faculty of Physics, University of Warsaw, 00-681 Warsaw, Poland}

\author{X. Liu}
\affiliation{Department of Physics, University of Notre Dame, Notre Dame, IN 46556, USA}

\author{J. K. Furdyna}
\affiliation{Department of Physics, University of Notre Dame, Notre Dame, IN 46556, USA}

\date{\today}

\begin{abstract}
Applicability of dilute magnetic semiconductors (DMS) in electronic devices relies upon the understanding and control of their magnetic anisotropy. This paper explores one of the ways in engineering magnetic anisotropy in epitaxial layers of DMS by forming them into magnonic structures. For this purpose the canonical ferromagnetic DMS, namely (Ga,Mn)As, is employed. The anti-dot systems based on (Ga,Mn)As layers of various thicknesses are fabricated with focused ion beam apparatus and studied by means of microscopy as well as magnetometry. The magnetometric data --- collected along two nonequivalent in-plane crystallographic directions of (Ga,Mn)As: $[110]$ and $[1\overline{1}0]$ --- shows the effect of structuring on high-field magnetization process, whereas no significant change of the width of hysteresis loop in anti-dot samples is observed.\end{abstract}


\maketitle

\section{\label{sec:intro}Introduction}
Ever since the fabrication of ferromagnetic layers of dilute magnetic semiconductors (DMS) was mastered,\cite{Munekata1989} it was suggested that these materials were suitable for applications in future electronic devices.\cite{Akinaga2002,Dietl2003} The success of DMS-based approach to electronics is dependent on two conditions: pushing Curie temperature ($T_{\mathrm{C}}$) of ferromagnet-to-paramagnet transition in DMS well above $300$ K and controllable engineering of magnetic anisotropy in such materials.

First of these tasks witnessed remarkable progress, mainly due to post-growth annealing of epitaxially grown DMS, with the highest observed Curie temperature being slightly below $200$ K.\cite{Wang2014,Zhao2013,Fukuma2008} Presently, one may notice some slowing down in pursuit of the above-room-temperature $T_{\mathrm{C}}$, whereas the methods other than annealing are exercised, e.g. harnessing proximity effects in hybrid structures involving DMS and robust metallic ferromagnets.\cite{Song2011}

The engineering of magnetic anisotropy in ferromagnetic DMS was preceded by long and extensive studies of native magnetic anisotropy in such materials and in (Ga,Mn)As specifically.\cite{Liu2006} Only very recently the puzzle of the uniaxial in-plane component of the magnetic anisotropy in (Ga,Mn)As has been solved, when the role of Mn-Mn pairs and uneven distribution of their orientations along $[110]$ or $[1\overline{1}0]$ directions has been elucidated.\cite{Birowska2012} One should emphasize here the importance of understanding and control of the uniaxial magnetic anisotropy: any functional electronic unit, which operation is based on switching the magnetization in bi-stable manner, is ultimately dependent on such $2$-fold symmetry of magnetic properties.

In order to move beyond the native features of magnetic anisotropy in DMS, different approaches were pursuit, including those employing strain or proximity effects.\cite{King2011,Dziatkowski2006} In this report yet another approach is proposed, the one applying magnonic crystal paradigm to DMS.\cite{Lenk2011} Magnonic crystals (or wider: magnonic structures) are the nano- and micro-engineered systems in which the periodic modification of magnetic properties induces specific, qualitatively new properties and phenomena,\cite{Kruglyak2010,Neusser2009} to mention only magnetic anisotropies of exotic symmetries,\cite{Crowburn2002,Tacchi2010} preferred propagation directions or localization of spin waves,\cite{Neusser2010,Hu2011} or frequency-selective suppression of magnons (so called magnonic gap).\cite{Wang2010,Costa2011} The aim of this study is to check for the feasibility of modification of the in-plane magnetic anisotropy in (Ga,Mn)As, and its uniaxial component in particular, by means of the magnonic structuring of epitaxially-grown DMS layers.

\section{\label{sec:samples}Samples and Methods}
\begin{figure}[!h]
\begin{center}
\includegraphics[width=\columnwidth]{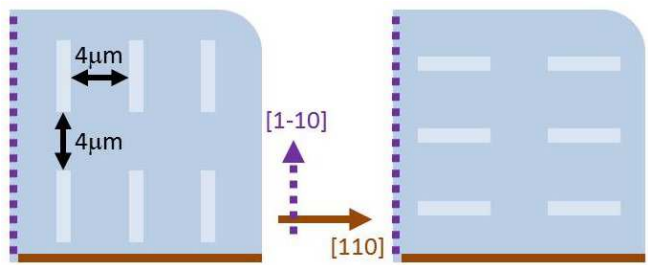}
\end{center}
\caption{\label{fig:trenches}Diagram of an anti-dot lattice differently oriented with respect to (Ga,Mn)As crystallographic axes.}
\end{figure}

Samples for this study were deposited on a $(001)$ GaAs semi-insulating substrate by means of molecular beam epitaxy (MBE). High temperature GaAs buffer was grown at $600$\textcelsius, whereas low temperature GaAs buffer and (Ga,Mn)As layer of interest were deposited at $250$\textcelsius\ in order to introduce about $5$\%\ of Mn into the latter film. No layer was annealed after the growth. This process resulted in spontaneous magnetization of the as-grown, unstructured layers in the range of $12$-$20$ emu/cm$^3$. Thickness $d$ of the ferromagnetic (Ga,Mn)As was varied between $12$ and $44$ nm. After the growth, a pair of rectangular samples being in the immediate proximity was cut from each layer. From now on in this paper the term ``layer'' will be used with respect to a product of MBE process having given thickness, while the term ``sample'' will refer to a piece cut from a given layer. Such a pair of samples was hereafter regarded as almost identical from the viewpoint of their magnetic properties and validity of this assumption was checked in magnetometric measurements. Then in each pair of samples the sets of periodically placed anti-dots in the form of $1$ $\mathrm{\mu}$m by $5$ $\mathrm{\mu}$m rectangular trenches were etched with focused ion beam (FIB) method, with longer edge of the rectangle along either $[110]$ or $[1\overline{1}0]$ axis of GaAs (see Fig. \ref{fig:trenches}).
Depth of the anti-dots was varied between $7$ and $12$ nm for (Ga,Mn)As layers of different thickness.

\begin{figure}[!h]
\begin{center}
\includegraphics[width=\columnwidth]{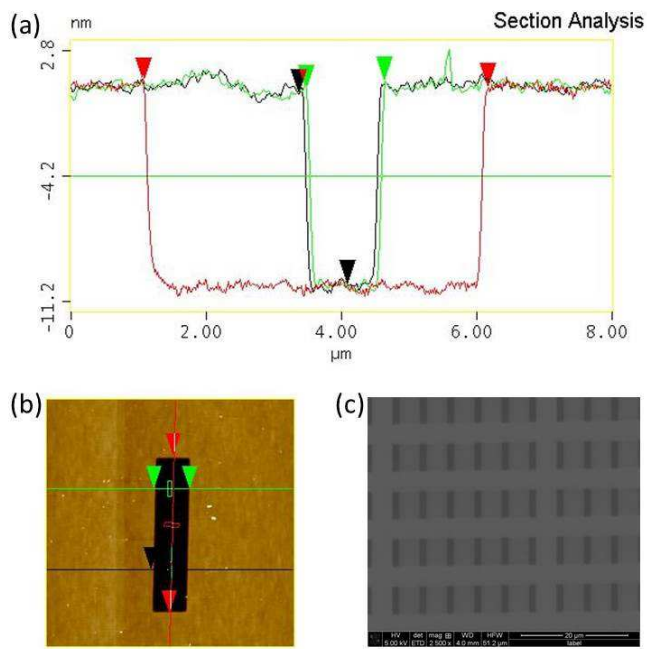}
\end{center}
\caption{\label{fig:sem_afm}(a)-(b) Scanning electron microscopy and (c) atomic force microscopy data for one of (Ga,Mn)As layers.}
\end{figure}

Samples with fabricated anti-dots were subsequently investigated by means of scanning electron microscopy (SEM) and atomic force microscopy (AFM). Typical results are shown in Fig. \ref{fig:sem_afm}.
The former method revealed good resemblance of etched lattices to the intended pattern, while the latter techniques brought the estimate for roughness of the trenches at about $1$ nm. One should also note that in few cases SEM data revealed some re-deposition of the etched material on the top of studied samples.

Finally the samples underwent magnetometric studies in a superconducting quantum interference device (SQUID). For the main purpose of this report the saturation curves were recorded, i.e. dependencies of magnetization $M$ on the applied field $H$ from $0$ to $70$ kOe. The measurements were carried out at room ($300$ K) and low ($5$ K) temperatures $T$, with the external magnetic field applied along either $[110]$ or $[1\overline{1}0]$ crystallographic directions, which are also the cleaving edges of GaAs. In addition to these data the hysteresis curves were collected in the narrower span of magnetic fields between $-10$ and $10$ kOe as well as temperature-dependent remnant magnetization curves were recorded at the very low field of $50$ Oe.

\section{\label{sec:results}Results and Discussion}
\begin{figure}[!h]
\begin{center}
\includegraphics[width=\columnwidth]{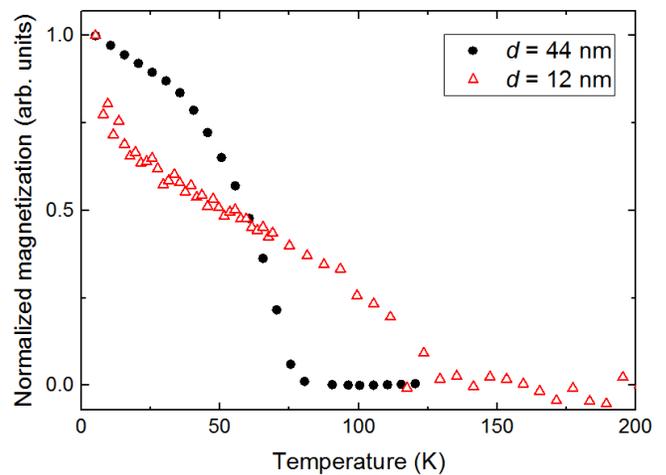}
\end{center}
\caption{\label{fig:temp}Spontaneous magnetization (divided by its value at the lowest temperature) for two (Ga,Mn)As layers.}
\end{figure}

(Ga,Mn)As layers fabricated for the purpose of this report were ferromagnetic, as evidenced by $M$-vs-$T$ data presented in Fig. \ref{fig:temp}. The observed Curie temperature of ferromagnet-to-paramagnet transition was in the range of $80$-$120$ K, depending on the layer. For the thickest layer of this study, $d=44$ nm, the Curie point is very well pronounced, confirming good control of the growth process. In the case of thinnest layer, $d=12$ nm, the collected data do not form the curve expected for the ferromagnet-to-paramagnet critical phenomena and the transition is much more diffused yet undoubtedly visible. Virtually zero magnetic moment seen in Fig. \ref{fig:temp} for temperatures above $120$ K suggests no significant contribution from MnAs clusters that remain ferromagnetic up to room temperature. This observation was corroborated by reflection high-energy electron diffraction patterns, which --- observed during MBE process --- confirmed good quality of deposited (Ga,Mn)As layers.

\begin{figure}[!h]
\begin{center}
\includegraphics[width=\columnwidth]{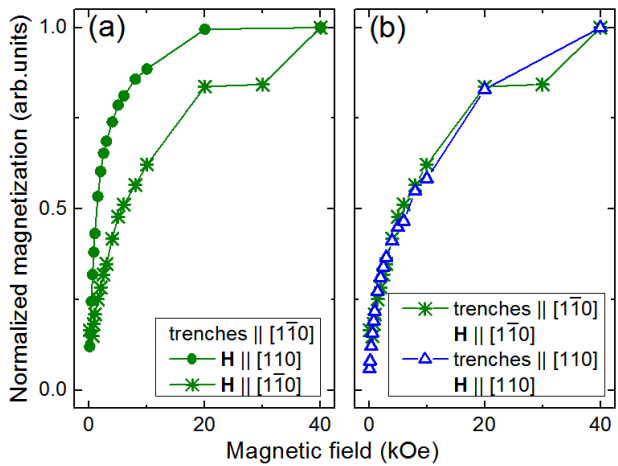}
\end{center}
\caption{\label{fig:hifield44}Magnetization at $300$ K with the external magnetic field applied to: (a) the same sample but with different orientations
with respect to anti-dots; (b) different samples but with the same orientation with respect to anti-dots; $d=44$ nm.}
\end{figure}

At room temperature, i.e. far above the Curie point of studied layers, no spontaneous magnetization of (Ga,Mn)As is expected although non-zero magnetic polarization can be induced by applying high external magnetic field. Also, there will be no effect of native magnetic anisotropy of (Ga,Mn)As at such conditions, thus any dependence on the direction of the applied field should be accredited to the influence of the anti-dot structuring. In Fig. \ref{fig:hifield44} the magnetization curves for one of the layers influenced by the elevated magnetic fields is presented. The data (collected from the initially magnetized state) were corrected for the unavoidable diamagnetic contribution from substrate and buffers, and then normalized to the saturation value, since at $H>40$ kOe the magnetic moment was roughly constant after correction.
The left panel shows two curves gathered for the same sample but for different orientations of the external field with respect to the anti-dots, whereas the right panel contains data collected from different samples of the same layer and at the same direction of $\mathbf{H}$ with respect to fabricated pattern. One can easily notice a dependence of the ``rate'' of magnetization process, i.e. of the slope of $M$-vs-$H$ curve, on the direction of $\mathbf{H}$ \emph{with respect to the anti-dots}. Apparently the anisotropic anti-dot pattern makes one of the in-plane directions in (Ga,Mn)As easier for high-field-induced magnetization than the other (the analogous effect was observed in other layers under study). Interestingly, out of four studied configurations of the external field and anti-dots, the one at which the magnetization curve saturates the quickest is with $\mathbf{H}$ perpendicular to the rectangular trenches, as shown in Fig. \ref{fig:hifield44}(a), although at first guess one could expect that shape anisotropy would prefer the direction along the anti-dots.

\begin{figure}[!h]
\begin{center}
\includegraphics[width=\columnwidth]{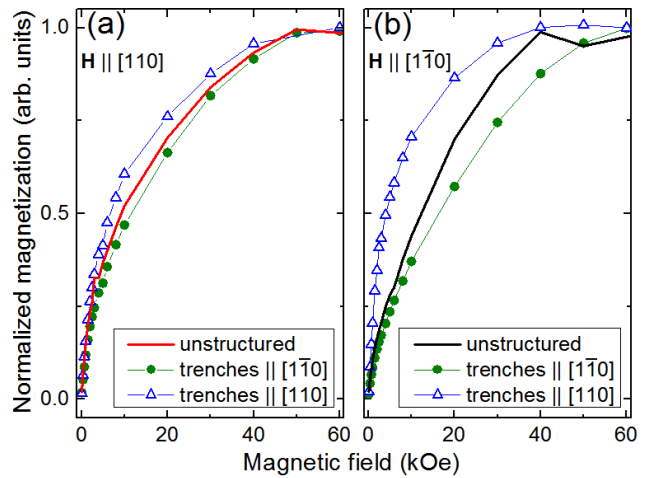}
\end{center}
\caption{\label{fig:hifield12}Magnetization curves collected at $5$ K for different orientations of the external magnetic field applied to $12$-nm-thick (Ga,Mn)As layer with and without anti-dots.}
\end{figure}

Figure \ref{fig:hifield12} presents $M$-vs-$H$ data collected (from the initially magnetized state) for $12$-nm-thick layer at $5$ K instead of room temperature, but the high field features of studied samples rendered to be qualitatively the same in both situations.
The differences, if any, should be expected at much smaller fields, when the component of magnetization free energy related to the external field is no longer dominant with respect to its anisotropic counterparts. These conditions will be discussed below. Most importantly Fig. \ref{fig:hifield12} presents a comparison between structured and unstructured samples and brings clear evidence for the effect of trenches on the ``rate'' of the magnetization process: for a given orientation of $\mathbf{H}$, the change in orientation of the trenches may increase or decrease the slope of $M$-vs-$H$ curve. Again the $M$-vs-$H$ curve saturates \emph{the quickest} for the external field applied perpendicular to the trenches (see panel (b) of Fig.\ref{fig:hifield12}), but this time for different orientation of $\mathbf{H}$ with respect to crystallographic axes. Moreover, as shown in the panel (a) of Fig.\ref{fig:hifield12}, the orientation of the external magnetic field at right angle to the trenches may render moderate decrease in the slope of $M$-vs-$H$ curve when compared with the unstructured sample. The explanation of such behavior requires detailed analysis of the spatial distribution of the internal magnetic field within a patterned samples, performed e.g. by means of micromagnetic simulations, which is beyond the scope of this report.

\begin{figure}[!h]
\begin{center}
\includegraphics[width=\columnwidth]{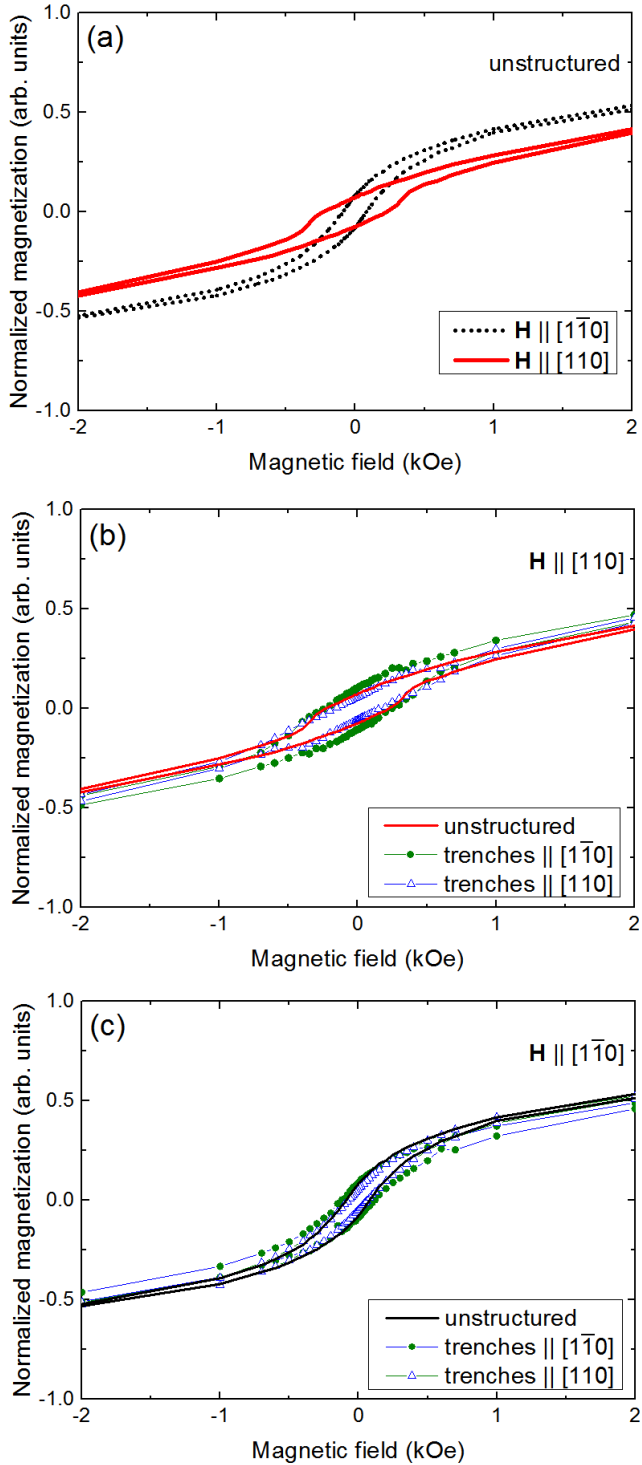}
\end{center}
\caption{\label{fig:hist30} (a) Hysteresis loops collected at $5$K for as-grown $30$-nm-thick layer of (Ga,Mn)As. (b)-(c) Comparison of hysteresis loops before and after anti-dot structuring.}
\end{figure}

Finally, the hysteresis loops collected at $5$ K for one of the layers under study are shown in Fig. \ref{fig:hist30}.
The unstructured samples revealed clear uniaxial character of in-plane magnetic anisotropy, with the $\mathbf{H}||[110]$ hysteresis loop being much wider than the one recorded with the external magnetic field applied along $[1\overline{1}0]$ crystallographic direction of (Ga,Mn)As. The hysteresis loops of samples being the subject to FIB process --- especially these presented in panel (b) of Fig. \ref{fig:hist30} --- seem to be more ovoid than their counterparts in panel (a). One may suspect that ion bombardment in FIB apparatus produces in (Ga,Mn)As layers some structural damage, which may be reflected in the quality of features observed in magnetic measurements. On the other hand the general uniaxial pattern of in-plane magnetic anisotropy is preserved in structured samples and there is no significant effect of trenches on the width of hysteresis loops visible in panels (b) and (c) of Fig. \ref{fig:hist30}. While the latter observation is regrettable from the viewpoint of this report aiming on the engineering of magnetic anisotropy in (Ga,Mn)As, it is also an indication that FIB-induced damages in studied samples, although unavoidable, were limited.

\section{\label{sec:conclusion}Conclusion}
The fabrication, characterization, and magnetometric study of the epitaxial (Ga,Mn)As layers with multiple anti-dot features were reported. The obtained FIB-etched structures form the simplest magnonic structure made of ferromagnetic dilute semiconductor, and were designed in order to modify the in-plane uniaxial magnetic anisotropy observed natively in (Ga,Mn)As. SEM and AFM data revealed good quality of fabricated trenches, with their proper alignment with respect to the crystallographic axes of (Ga,Mn)As and only limited re-deposition of the etched material. Subsequent SQUID magnetometry studies showed that anti-dot structure clearly affects the field-induced magnetization process in (Ga,Mn)As, what was manifested by the change in slope of $M$-vs-$H$ curves collected at high external fields. At the lowest magnetic fields applied, no significant dependence of the width of the hysteresis loop on the anti-dot pattern was observed.

\begin{acknowledgments}
This work was supported by the European Union within the European Regional Development Fund through the Innovative Economy program under the grant Homing Plus/2011-4/6 from the Foundation for Polish Science.
\end{acknowledgments}


\begin{thebibliography}{99}
\bibitem{Munekata1989} H. Munekata, H. Ohno, S. von Molnar, A. Segmuller, L. L. Chang, and L. Esaki, "Diluted magnetic III-V semiconductors", \emph{Phys. Rev. Lett.}, vol. 63, no. 17, pp. 1849-1852, Oct. 1989.
\bibitem{Akinaga2002} H. Akinaga and H. Ohno, "Semiconductor spintronics", \emph{IEEE Trans. Nanotech.}, vol. 1, no. 1, pp. 19-31, Mar. 2002.
\bibitem{Dietl2003} T. Dietl, "Dilute magnetic semiconductors: Functional ferromagnets", \emph{Nature Mat.}, vol. 2, no. 10, pp. 646-648, Oct. 2003.
\bibitem{Wang2014} M. Wang, R. A. Marshall, K. W. Edmonds, A. W. Rushforth, R. P. Campion, and B. L. Gallagher, "Determining Curie temperatures in dilute ferromagnetic semiconductors: High Curie temperature (Ga,Mn)As", \emph{Appl. Phys. Lett.}, vol. 104, no. 13, p. 132406, Apr. 2014.
\bibitem{Zhao2013} K. Zhao, Z. Deng, X. C. Wang, W. Han, J. L. Zhu, X. Li, Q. Q. Liu, R. C. Yu, T. Goko, B. Frandsen, L. Liu, F. L. Ning, Y. J. Uemura, H. Dabkowska, G. M. Luke, H. Luetkens, E. Morenzoni, S. R. Dunsiger, A. Senyshyn, P. B\"{o}ni, and C. Q. Jin, "New diluted ferromagnetic semiconductor with Curie temperature up to $180$ K and isostructural to the '122` iron-based superconductors", \emph{Nature Comm.}, vol. 4, p. 1442, Feb. 2013.
\bibitem{Fukuma2008} Y. Fukuma, H. Asada, S. Miyawaki, T. Koyanagi, S. Senba, K. Goto, and H. Sato, "Carrier-induced ferromagnetism in Ge$_{0.92}$Mn$_{0.08}$Te epilayers with a Curie temperature up to $190$ K", \emph{Appl. Phys. Lett.}, vol. 93, no. 25, p. 252502, Dec. 2008.
\bibitem{Song2011} C. Song, M. Sperl, M. Utz, M. Ciorga, G. Woltersdorf, D. Schuh, D. Bougeard, C. H. Back, and D. Weiss, "Proximity Induced Enhancement of the Curie Temperature in Hybrid Spin Injection Devices", \emph{Phys. Rev. Lett.}, vol. 107, no. 5, p. 056601, July 2011.
\bibitem{Liu2006} X. Liu and J. K. Furdyna, "Ferromagnetic resonance in Ga$_{1-x}$Mn$_{x}$As dilute magnetic semiconductors", \emph{J. Phys.: Condens. Matter}, vol. 18, no. 13, p. R245, Apr. 2006.
\bibitem{Birowska2012} M. Birowska, C. \'{S}liwa, J. Majewski, and T. Dietl, "Origin of Bulk Uniaxial Anisotropy in Zinc-Blende Dilute Magnetic Semiconductors", \emph{Phys. Rev. Lett.}, vol. 108, no. 23, p. 237203, June 2012.
\bibitem{King2011} C. S. King, J. Zemen, K. Olejn\'{i}k, L. Hor\'{a}k, J. A. Haigh, V. Nov\'{a}k, A. Irvine, J. Ku\v{c}era, V. Hol\'{y}, R. P. Campion, B. L. Gallagher, and T. Jungwirth, "Strain control of magnetic anisotropy in (Ga,Mn)As microbars", \emph{Phys. Rev. B}, vol. 83, no. 11, p. 115312, Mar. 2011.
\bibitem{Dziatkowski2006} K. Dziatkowski, Z. Ge, X. Liu, and J. K. Furdyna, "Identification of unidirectional anisotropy in exchange-biased MnO/GaMnAs bilayers using ferromagnetic resonance", \emph{Appl. Phys. Lett.}, vol. 88, no. 14, p. 142513, Apr. 2006.
\bibitem{Lenk2011} B. Lenk, H. Ulrichs, F. Garbs, and M. M\"{u}nzenberg, "The building blocks of magnonics", \emph{Phys. Rep.}, vol. 507, no. 4-5, pp. 107-136, Oct. 2011.
\bibitem{Kruglyak2010} V. V. Kruglyak, S. O. Demokritov, and D. Grundler, "Magnonics", \emph{J. Phys. D: Appl. Phys.}, vol. 43, no. 26, p. 264001, June 2010.
\bibitem{Neusser2009} S. Neusser and D. Grundler, "Magnonics: Spin Waves on the Nanoscale", \emph{Adv. Mater.}, vol. 21, no. 28, pp. 2927-2932, July 2009.
\bibitem{Crowburn2002} R. P. Crowburn, "Magnetic nanodots for device applications", \emph{J. Magn. Magn. Mater.}, vol. 242-245, no. 1, pp. 505-511, Apr. 2002.
\bibitem{Tacchi2010} S. Tacchi, M. Madami, G. Gubbiotti, G. Carlotti, H. Tanigawa, T. Ono, and M. P. Kostylev, "Anisotropic dynamical coupling for propagating collective modes in a two-dimensional magnonic crystal consisting of interacting squared nanodots", \emph{Phys. Rev. B}, vol. 82, no. 2, p. 024401, July 2010.
\bibitem{Neusser2010} S. Neusser, G. Duerr, H. G. Bauer, S. Tacchi, M. Madami, G. Woltersdorf, G. Gubbiotti, C. H. Back, and D. Grundler, "Anisotropic Propagation and Damping of Spin Waves in a Nanopatterned Antidot Lattice", \emph{Phys. Rev. Lett.}, vol. 105, no. 6, p. 067208, Aug. 2010.
\bibitem{Hu2011} C.-L. Hu, R. Magaraggia, H.-Y. Yuan, C. S. Chang, M. Kostylev, D. Tripathy, A. O. Adeyeye, and R. L. Stamps, "Field tunable localization of spin waves in antidot arrays", \emph{Appl. Phys. Lett.}, vol. 98, no. 26, p. 262508, July 2011.
\bibitem{Wang2010} Z. K. Wang, V. L. Zhang, H. S. Lim, S. C. Ng, M. H. Kuok, S. Jain, and A. O. Adeyeye, "Nanostructured Magnonic Crystals with Size-Tunable Bandgaps", \emph{ACS Nano}, vol. 4, no. 2, pp. 643-648, Feb. 2010.
\bibitem{Costa2011} C. H. O. Costa, M. S. Vasconcelos, and E. L. Albuquerque, "Partial band gaps in magnonic crystals", \emph{J. Appl. Phys.}, vol. 109, no. 7, p. 07D319, Mar. 2011.
\end{thebibliography}
\end{document}